\documentstyle[preprint,aps,psfig,amsmath]{revtex}

\begin{document}

\begin{center} 
Insensitivity of the Yrast Spectra of Even-Even Nuclei to the T=0 two-body interaction matrix elements
\medskip 
\\ 
Shadow J.Q. Robinson and Larry Zamick

Department of Physics and Astronomy,
Rutgers University, Piscataway, \\New Jersey  08855

\end{center}

\bigskip
\begin{abstract}
Calculations of the spectra of various even-even nuclei in the fp
shell ($^{44}$Ti, $^{46}$Ti, $^{48}$Cr, and $^{50}$Cr) are performed
with two sets of two-body interaction matrix elements. The first set
consists of the matrix elements of the FPD6 interaction. The second
set have the same T=1 two-body matrix elements as the FPD6
interaction, but all the T=0 two-body matrix elements are set equal to
zero. Despite the drastic differences between the two interactions,
the spectra they yield are very similar and indeed it is difficult to
say which set gives a better fit to experiment. That the results for
the yrast spectra are insensitive to the presence or absence of T=0
two-body matrix elements is surprising because the only bound two
nucleon system has T=0, namely the deuteron.  Also there is the
general folklore that T=0 matrix elements are responsible for nuclear
collectivity. Electric quadrupole transition rates are also
examined. It is found that the reintroduction of T=0 matrix elements
leads to an enhancement of B(E2)'s for lower spin transitions but in
some cases for higher spin transitions one gets another surprising
result that there is a small suppression.
\end{abstract}
\vspace{2.5in}

\newpage

\section{Introduction}

The study of neutron-proton pairing especially in the T=0 channel is a
particularly prominent topic these days. While the number of journal
articles are far too numerous to reference one might begin to make
some headway into the varied approaches by starting from the
references found in Refs.~\cite{goodmannew,macc1,8488}. In so doing
one will find a field of study filled with disagreement and occasionally
strife.

The approach here is to study the effects of the T=0 portion of the
neutron-proton interaction by removing it altogether. In this way it
is hoped that we can understand where in nuclear structure this
portion of the interaction is found to play a large role. This would
in turn suggest where T=0 pairing might most clearly reveal its
presence or absence. In this work we will examine the yrast spectra of
the even-even $(fp)^4$ nuclei $^{44}$Ti,$^{46}$Ti, $^{48}$Cr, and
$^{50}Cr$ with two sets of interaction parameters. First we have the
full FPD6 interaction~\cite{wrichter1}, and then we use the same
interaction for the two-body T=1 matrix elements while setting all the
T=0 two-body interaction matrix elements to zero. We shall denote this
interaction as T0FPD6. (This modification of an effective interaction
is along the same lines of that used by Satula et al. to examine
Wigner energies a few years ago~\cite{satula1}.)  We have used this
modification of FPD6 in the past to study a variety of things and in
particular the full fp spectrum of
$^{44}$Ti~\cite{dow1,dow2,dow3,dow4}. We found to our surprise and to
the surprise of many others that one could obtain a fairly decent
spectrum for the levels of $^{44}$Ti with the second interaction here
labeled T0FPD6. This is surprising because the T=0 interaction is by
no means small. The only bound two nucleon system is the T=0
combination of two nucleons, the deuteron. Moreover the largest valued
matrix elements in the FPD6 interaction tend to be those in the T=0
channel.

With the one nucleus $^{44}$Ti shown in a full fp space calculation in
Ref.~\cite{dow1} we have the first clue that the spectra of even-even
nuclei are relatively insensitive to the T=0 two-body interaction
matrix elements. To make this more conclusive however we need to
examine more nuclei. We have expanded the examination of the even-even
nuclei to those listed above. The sample we have chosen consists of
nuclei with the same collective properties - ground state bands which
have some rotational properties but are not extremely rotational.

Also we now look at transition rates - perhaps these will prove more
sensitive to the presence of T=0 two-body matrix elements than the
spectra. This will be our interest in a few sections.

It should be noted that in Refs.~\cite{dow1,dow2,dow3,dow4} a wide
range of topics is addressed beyond the spectra of even-even
nuclei. These topics include a partial dynamical symmetry that arises
when one uses the T0FPD6 interaction in a single j shell for $^{43}$Sc
and $^{44}$Ti. Also while using the T0FPD6 interaction a subtle
relationship between the T=$\frac{1}{2}$ states in $^{43}$Sc and
T=$\frac{3}{2}$ states in $^{43}$Ca likewise between the the T=$0$
states in $^{44}$Ti and T=$2$ states in $^{44}$Ca.  We also considered
even-odd nuclei and addressed the topic of how the T=0 two-body matrix
elements affect B(M1) transitions - both spin and orbital components,
and Gamow-Teller transitions. In many cases the transition rates were
very sensitive to the presence or absence of the T=0 matrix
elements. This was especially the case for some orbital B(M1)'s and
the Gamow-Teller transitions.

Here things will be kept simple and we focus on the spectra and
B(E2)'s of the yrast levels of selected even-even nuclei.  We will
examine the sensitivity of these observables on the T=0 two-body
interaction matrix elements by setting them to zero and comparing the
results thus obtained with those when the T=0 matrix elements are
reintroduced.

\section{Results}

\begin{figure}
\psfig{file=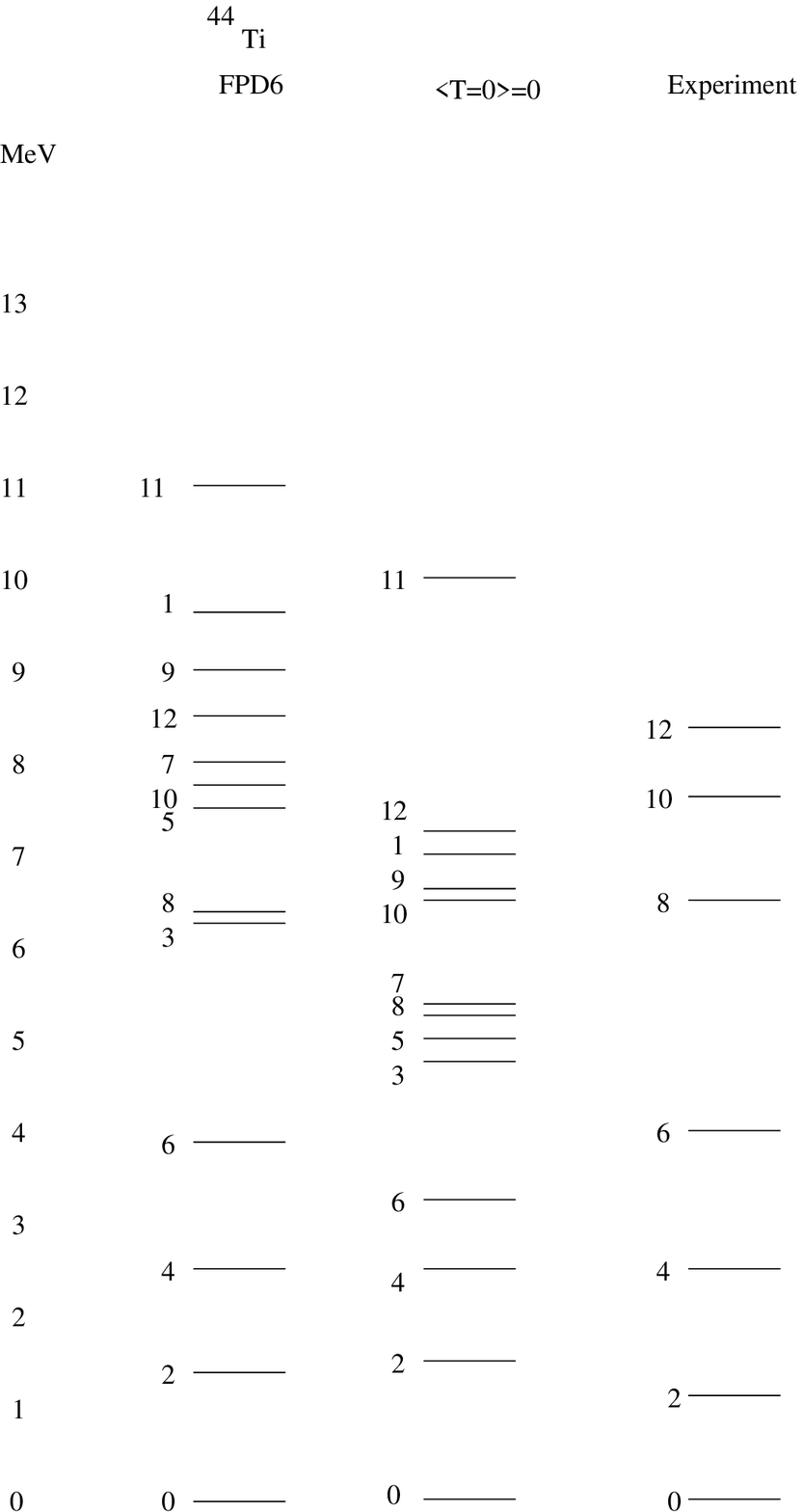,height=8in}
\caption{Full fp calculation for T=0 states in $^{44}$Ti with lowest state at 0 MeV.}
\label{fig:first}
\end{figure}

\begin{figure}
\psfig{file=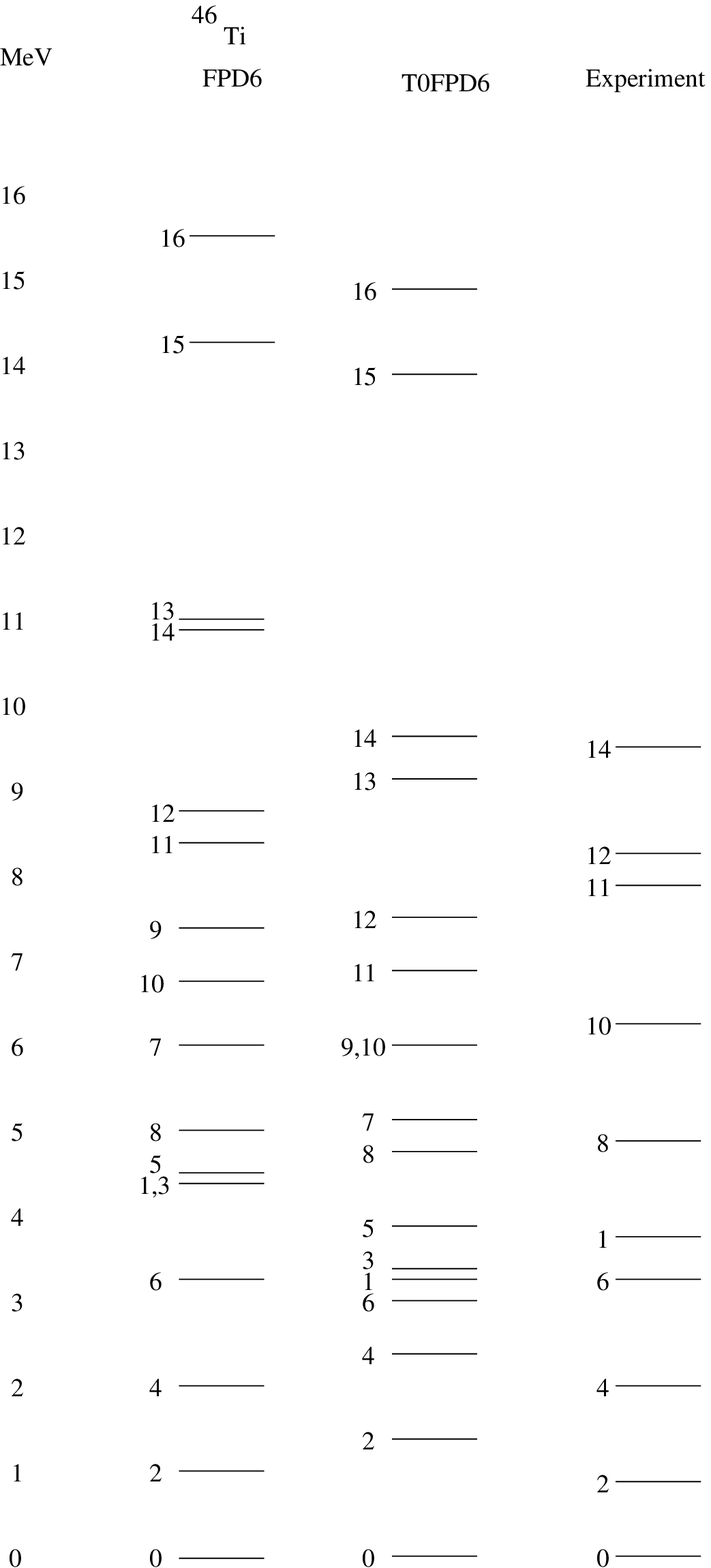,height=8in}
\caption{Full fp calculation for T=1 states in $^{46}$Ti with lowest state at 0 MeV.}
\label{fig:second}
\end{figure}

\begin{figure}
\psfig{file=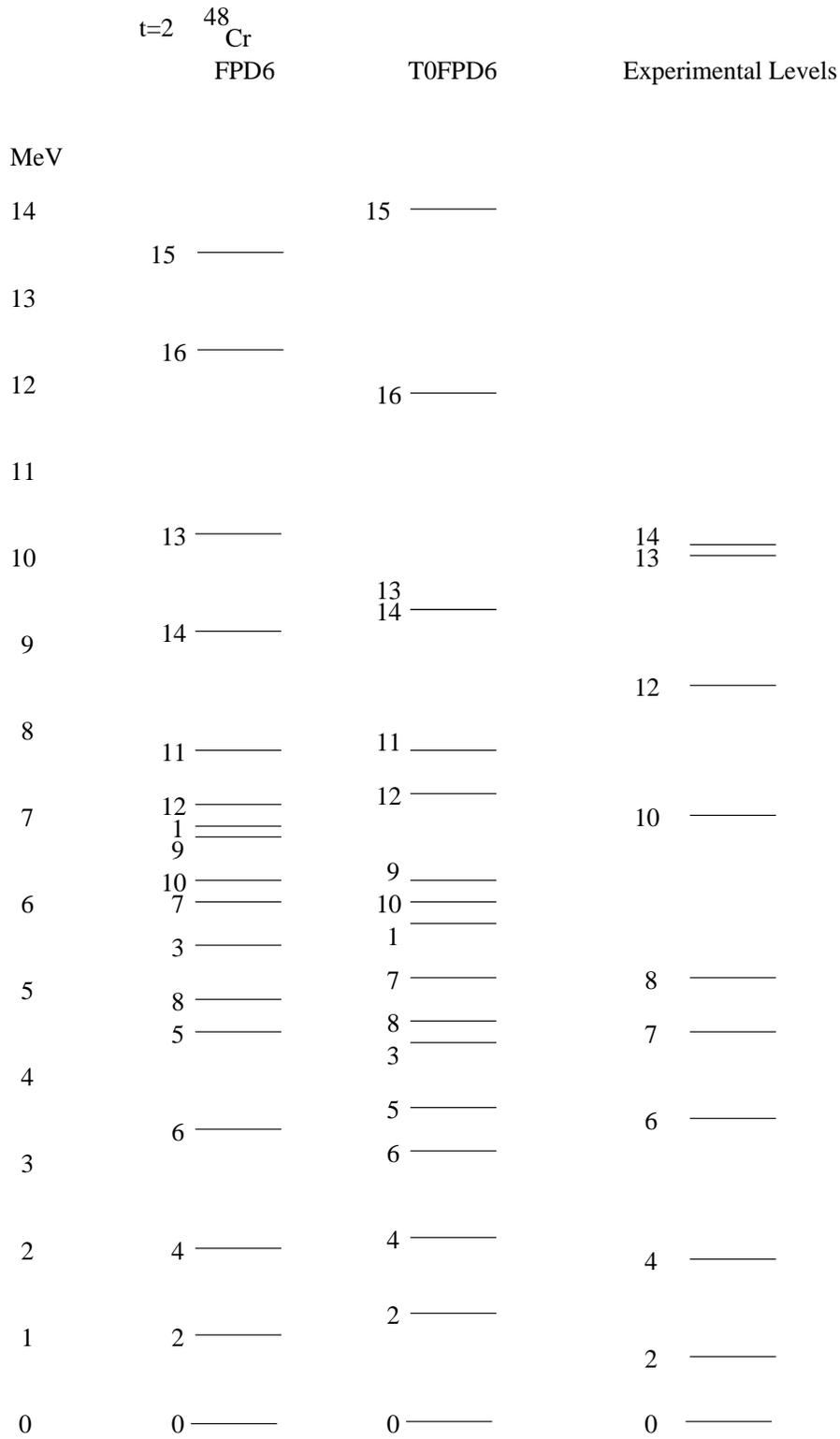,height=8in}
\caption{t=2 calculation and experimental results 
for T=0 states in $^{48}$Cr up to J=16}
\label{fig:third}
\end{figure}

\begin{figure}
\psfig{file=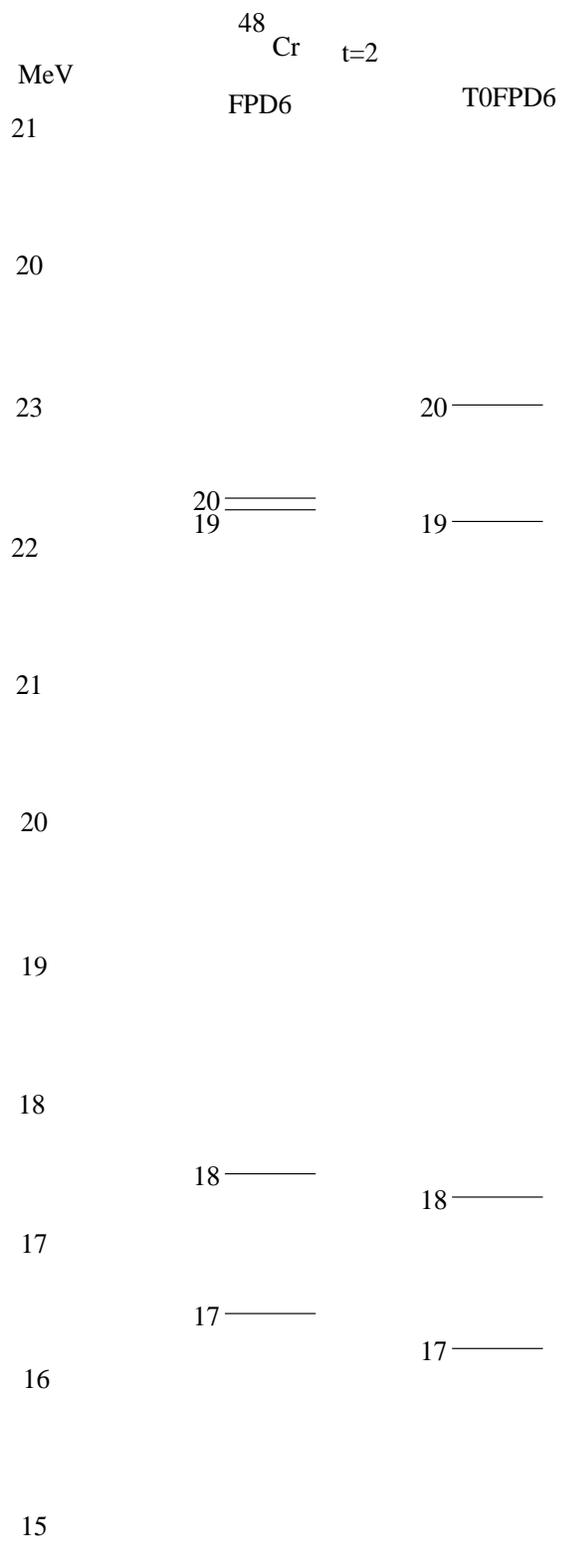,height=8in}
\caption{t=2 calculation for T=0 states in $^{48}$Cr J=17 and greater.}
\label{fig:fourth}
\end{figure}

\begin{figure}
\psfig{file=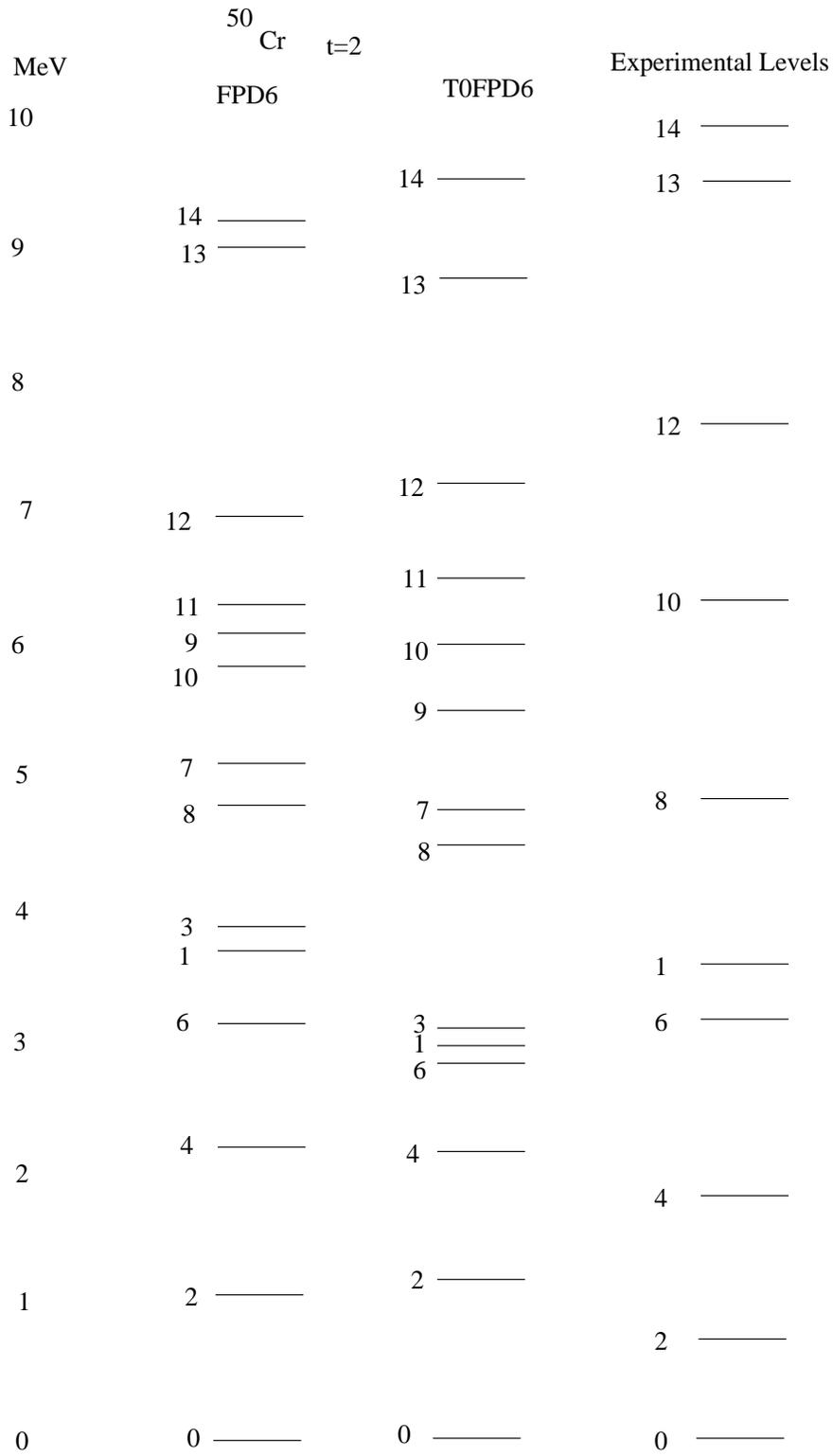,height=8in}
\caption{t=2 calculation and experimental results 
for T=1 states in $^{50}$Cr up to J=14}
\label{fig:fifth}
\end{figure}

\begin{figure}
\psfig{file=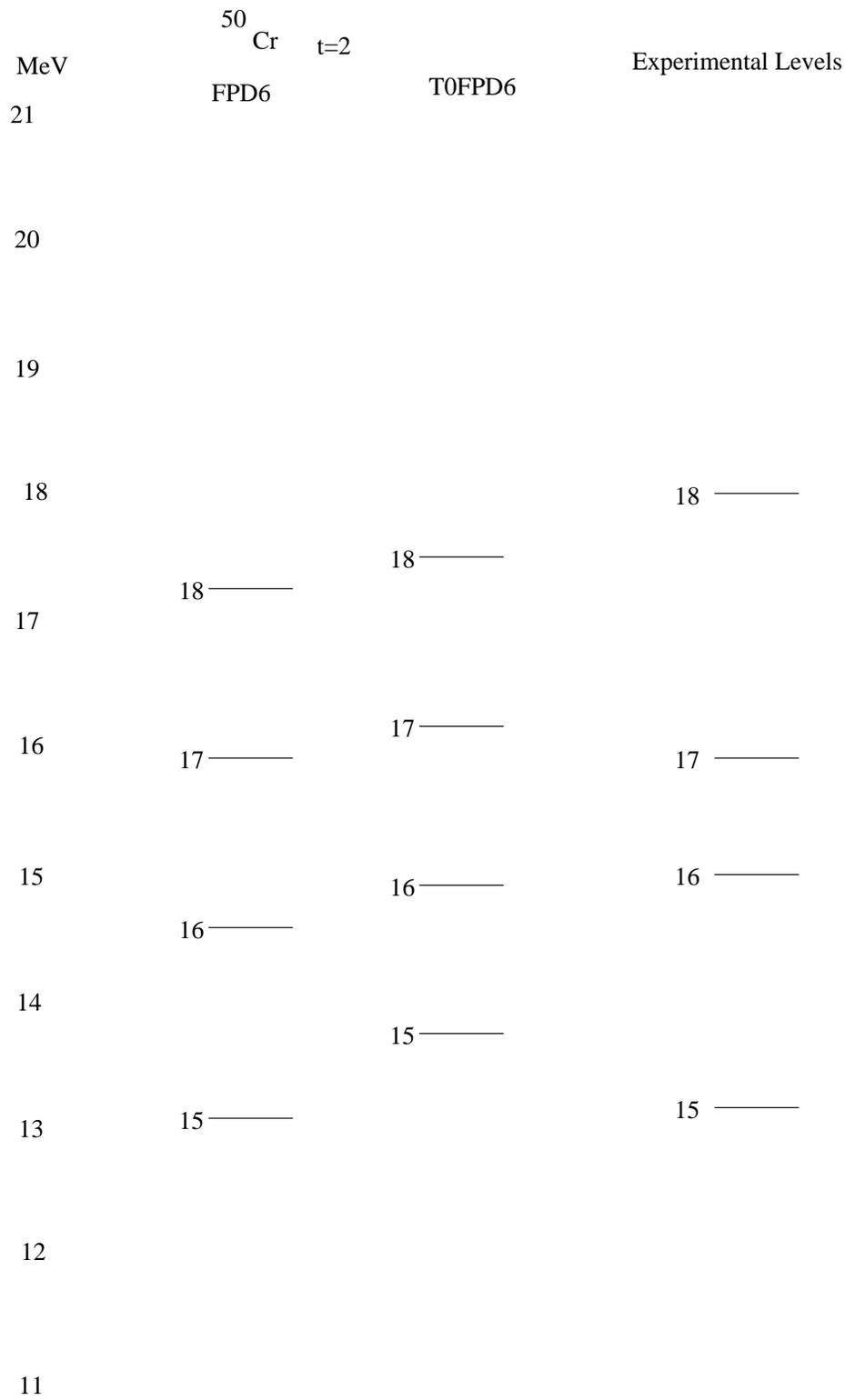,height=8in}
\caption{t=2 calculation for T=1 states in $^{50}$Cr J=15 and greater.}
\label{fig:last}
\end{figure}

As mentioned in the introductions we perform calculations of even-even
nuclei with and without the T=0 two-body matrix elements of the FPD6
interaction. Thus each of the figures \ref{fig:first} to
\ref{fig:last} corresponding to the nuclei $^{44}$Ti, $^{46}$Ti,
$^{48}$Cr, and $^{50}$Cr will consist of three columns.  The first
column is the Yrast spectra calculated with the full FPD6 interaction,
the second column with the T0FPD6 interaction, and the third column
shows the levels from experiment.

We have previously discussed $^{44}$Ti so we will start with $^{46}$Ti,
but before dissecting the details of the spectra, note the exceptional
results in column 2 of figure \ref{fig:second}. There we see that the
results found when all the T=0 two-body interaction matrix elements
are set to zero (T0FPD6) in this complete fp calculation that the
resulting spectra looks quite reasonable in comparison with both the
full FPD6 and the known experimental levels.  Indeed it is difficult
to choose between T0FPD6 and FPD6 as to which yields a better fit to
experiment.

A closer look shows some differences. The odd spin excitation
energies come down by about 1 MeV when the T=0 matrix elements are set
to zero. Experimentally some odd spin excitation energies are known
(J=1$^{+}$ and J=11$^{+}$) but not as many as the even spins.  The
even spin spectrum is slightly more spread out in the full FPD6
calculation that is to say there is more of a tendency towards a
rotational spectrum. The low lying spectrum is fit somewhat better
with the full FPD6 interaction but there is a better fit to the
J=14$^+$ state using the T0FPD6 interaction.  

There are similar stories for $^{48}$Cr (Figures
\ref{fig:third},\ref{fig:fourth}) and $^{50}$Cr (Figures
\ref{fig:fifth} and \ref{fig:last}). (These nuclei are of some
interest as they both display
backbends~\cite{lenzi1}-~\cite{burrows2}.) In $^{44}$Ti there is a bigger
difference between FPD6 and T0FPD6 perhaps results from the greater
deformation in $^{44}$Ti.

As is the case for $^{46}$Ti and indeed many of the even-even nuclei
in the region very few odd spin states are known.  If these were to be
made available experimentally it might be easier to demonstrate more
clearly the preference between the full FPD6 or the modification
T0FPD6. The fact that one can even think of offering into
competition an interaction in which all the T=0 two-body interaction
matrix elements are set to zero is quite remarkable and perhaps a bit
disturbing.

\section{B(E2) rates}

The B(E2) rates for $^{44}$Ti, $^{46}$Ti, $^{48}$Cr, and $^{50}$Cr are
listed in tables \ref{tab:first} to \ref{tab:last}. We allow up to t
nucleons to be excited from the f$_{7/2}$ shell to the rest of the fp
shell. The values of t used are 4,3,2, and 2 respectively. The
effective charges used are the standard $1.5e$ for the proton and $0.5
e$ for the neutron. The difference in the effective charges from 1 and
0 is intended to take care of the fact that the $\Delta N =2$ and
higher excitations are not present in this model space.  The results
for FPD6 and T0FPD6 are shown. We also display the ratios of the
results for the 2 interactions.

For $^{46}$Ti the reintroduction of the T=0 two-body matrix elements
causes an increase (relative to T0FPD6) of a factor of two or more
for all the transitions considered. So there is evidence here that the
T=0 matrix elements contribute to the collectivity. This is not seen
by looking at just the excitation energies in $^{46}$Ti where FPD6 and
T0FPD6 give very similar results.

In $^{48}$Cr and $^{50}$Cr a more interesting behavior evolves. In t=2
calculations, the B(E2)'s for low spin transitions get and enhancement
with the reintroduction of the T=0 two-body matrix elements but for
some of the higher spins we get a suppression. The suppression occurs
when backbending occurs as in $^{50}$Cr in the J=8 $\rightarrow$ 10
transition. 

In summary, in studying the problem of the T=0 neutron-proton
interaction in a nucleus, it may prove more fruitful to begin by
removing this channel altogether as was done here by setting all the
T=0 two-body matrix elements elements to zero and then reintroducing
them, rather than adopting the more common approach of investigating
the effects of a pairing interaction separated from the rest of the
interaction.  This may be especially true in the shell model as the
suggestion has been made by Satula and Wyss that it may not be
appropriate to separate out a pairing interaction from the rest of the
Hamiltonian in a shell model context~\cite{satw676}.

This work was supported by the U.S. Dept. of Energy under Grant
No. DE-FG02-95ER-40940 and one of us by GK-12 NSF9979491
(SJQR).

\begin{table}
\begin{center}
\caption{$^{44}$Ti yrast B(E2) values ($e^{2}fm^{4}$) in FPD6 
and  T0FPD6} 
\label{tab:first}
\begin{tabular}{cccc}
                  & t=3 & t=3 &ratio \\
                  & FPD6&T0FPD6 & \\
0 $\rightarrow$ 2  &702.2&433.2 &0.617 \\
2 $\rightarrow$ 4  &344.3&169.1 &0.491  \\
4 $\rightarrow$ 6  &233.8&70.85 &0.303 \\
6 $\rightarrow$ 8  &147.8&75.64 &0.512 \\
8 $\rightarrow$ 10 &135.6&90.72 &0.669 \\
10 $\rightarrow$ 12&75.65&55.65 &0.736 \\
\end{tabular}
\end{center}
\end{table}

\begin{table}
\begin{center}
\caption{$^{46}$Ti yrast B(E2) values ($e^{2}fm^{4}$) in FPD6 
and  T0FPD6} 
\label{tab:second}
\begin{tabular}{cccc}
                  & t=3 & t=3 &ratio \\
                  & FPD6&T0FPD6 & \\
0 $\rightarrow$ 2  &672.1&472.6& 0.703\\
2 $\rightarrow$ 4  &N/A  &N/A  & N/A \\
4 $\rightarrow$ 6  &272.1&100.8& 0.370\\
6 $\rightarrow$ 8  &221.6&93.13& 0.420\\
8 $\rightarrow$ 10 &169.4&84.84& 0.501\\
10 $\rightarrow$ 12&63.64&33.45& 0.526\\
12 $\rightarrow$ 14&46.13&22.07& 0.478\\
14 $\rightarrow$ 16&1.407&0.5186& 0.369\\
\end{tabular}
\end{center}
\end{table}

\begin{table}
\begin{center}
\caption{$^{48}$Cr yrast B(E2) values ($e^{2}fm^{4}$) in FPD6
 and T0FPD6} 
\label{tab:third}
\begin{tabular}{cccc}
&t=2 &t=2& ratio \\ 
&FPD6& T0FPD6&  \\
0 $\rightarrow$ 2  & 892   &691.6 &0.775\\
2 $\rightarrow$ 4  & 424.7 &287.8 &0.678\\
4 $\rightarrow$ 6  & 345   &169.4 &0.491\\
6 $\rightarrow$ 8  & 319.5 &202.1 &0.633\\
8 $\rightarrow$ 10 & 238.9 &175.9 &0.736\\
10 $\rightarrow$ 12& 168.9 &133   &0.787\\
12 $\rightarrow$ 14& 138   &120.7 &0.875\\
14 $\rightarrow$ 16& 77.60 &79.22 &1.021\\
16 $\rightarrow$ 18& 1.178 &1.471 &1.249\\
18 $\rightarrow$ 20& 2.555 &0.728 &0.285\\
\end{tabular}
\end{center}
\end{table}

\begin{table}
\begin{center}
\caption{$^{50}$Cr yrast B(E2) values ($e^{2}fm^{4}$) in FPD6
 and T0FPD6} 
\label{tab:last}
\begin{tabular}{cccc}
&t=2 &t=2 & ratio \\ 
&FPD6& T0FPD6&  \\
0 $\rightarrow$ 2  & 761.8 & 614.6 & 0.807\\
2 $\rightarrow$ 4  & N/A   & N/A   & N/A \\
4 $\rightarrow$ 6  & N/A   & N/A   & N/A\\
6 $\rightarrow$ 8  & 188.3 & 133   & 0.706\\
8 $\rightarrow$ 10 & 46.48 & 69.12 & 1.486\\
10 $\rightarrow$ 12& 54.83 & 63.79 & 1.163\\
12 $\rightarrow$ 14& 74.58 & 80.91 & 1.085\\
14 $\rightarrow$ 16& 6.665 & 3.647 & 0.547\\
16 $\rightarrow$ 18& 86.84 & 36.68 & 0.422\\
18 $\rightarrow$ 20& 0.8343& 1.006 & 1.206\\
\end{tabular}
\end{center}
\end{table}

\end{document}